# On the "unreasonable" effectiveness of Transport of Intensity imaging and optical deconvolution

T.E. Gureyev[1,2,3,4,*], Ya.I. Nesterets[4,3], A. Kozlov[1], D.M. Paganin[2], and H.M. Quiney[1]

[1]ARC Centre of Excellence in Advanced Molecular Imaging, The University of Melbourne, Parkville, VIC 3010, Australia
[2]School of Physics and Astronomy, Monash University, Clayton, VIC 3800, Australia
[3]University of New England, Armidale, NSW 2351, Australia
[4]Commonwealth Scientific and Industrial Research Organisation, Clayton, VIC 3168, Australia
[*]timur.gureyev@unimelb.edu.au

9 August 2017

**Abstract**

The effectiveness of reconstructive imaging using the Homogeneous Transport of Intensity equation may be regarded as "unreasonable", because it has been shown to significantly increase signal-to-noise ratio while preserving spatial resolution, compared to equivalent conventional absorption-based imaging techniques at the same photon fluence. We reconcile this surprising behaviour by analysing the propagation of noise in typical in-line holography experiments. This analysis indicates that novel imaging techniques may be designed which produce high signal-to-noise images at low radiation doses without sacrificing spatial resolution.

## 1. Introduction

It can be shown, using elementary arguments, that the spatial resolution and signal-to-noise ratio (SNR) in free-space imaging (or scattering) experiments are limited primarily by the number of incident photons that are made to interact with (scattered by) each "voxel" of the sample (see e.g. [1-5]). In that context, it is almost self-evident that, if the incident photon fluence remains fixed, any improvement in the spatial resolution (i.e. any reduction in the size of the resolvable elemental interacting volume) can only come at the expense of SNR (as there are proportionally fewer photons scattered by the smaller volume on average) and vice versa. This leads to the so-called noise-resolution uncertainty principle that has been proven in the context of optical imaging [5,6] and, more recently, in quantum electrodynamics [7]. In the latter context it has been shown [7] that this principle is inherent to quantum field theory and can even be used to refine the lower bound in the Heisenberg Uncertainty Principle in certain circumstances.

Given this fundamental nature of the trade-off between the SNR and spatial resolution in any imaging or scattering experiments, it is very surprising that the reconstructive imaging method based on the so-called Homogeneous Transport of Intensity equation (TIE-Hom) [8] is able to comprehensively break the limit set by the noise-resolution uncertainty principle and increase SNR in the reconstructed images, sometimes by a factor of hundreds, without sacrificing spatial resolution [9,10]. One of the goals of the present work is to find a theoretical explanation for this apparent paradox. It is important to note that two orders of magnitude increase in SNR enables four orders of magnitude decrease in acquisition time and associated dose, with obvious positive ramifications in both biomedical and industrial imaging contexts. Notwithstanding the uptake of the TIE-Hom method in several hundred papers spanning science, engineering and medicine, particularly when combined with tomographic reconstruction, its hitherto-mysterious SNR boost has never been satisfactorily explained in the existing literature. It is timely that such an explanation of the "unreasonable" effectiveness of the TIE-Hom method be provided.

We show that in in-line imaging, the behaviour of image noise is in some ways parallel to that of spatial resolution. It is well known that the square of the spatial resolution of an in-line imaging system, defined as the variance of the point-spread function (PSF), is equal to a weighted sum of the variances of the source intensity distribution and the detector PSF, with weights determined only by the geometric magnification inherent to the imaging setup [11]. Under some rather general imaging conditions considered in the present paper, the variance of the image noise is similarly a weighted sum of the source-attributed noise variance (typically satisfying Gaussian statistics) and the photodetection (shot) noise variance (typically, with Poisson statistics). Both terms of this linear combination are weighted in accordance with certain geometric factors, which are similar to those for the spatial resolution. This "parallel" behaviour is closely related to the known invariance of the ratio of the SNR and spatial resolution under linear shift-invariant transformations of imaging systems [5]. Careful observation of the different behaviours of the two components of image noise allows us to find a quantitative explanation for the "unreasonable" effectiveness of TIE-Hom imaging. The key to this explanation is the behaviour of the transverse correlation length of the noise during forward free-space propagation and in numerical TIE-Hom retrieval. The details of the explanation of the TIE-Hom "paradox" can be found in Section 3 below, after the theory of free-space propagation of intensity covariance is studied in Section 2. Many results presented in Section 2 can be found in a slightly different form in Ref. 12.

Given the theoretical model developed in the present paper, it is conceivable that the favourable properties of TIE-Hom imaging can be utilised for devising novel optical methods with similar properties. An indication of possible generalisations of the key favourable properties of TIE-Hom imaging to other forms of imaging system is provided at the end of the present paper. Such methods would be also able to produce high-SNR and high-resolution images at low photon levels. It is likely that such methods could be very useful in areas of imaging where "information-carrying" photons are scarce (as in some areas of astronomy), or where the samples are sensitive to the radiation dose (as in medical and biomedical imaging and microscopy), or where high throughput makes reduced acquisition time of high importance (as in high-volume manufacturing quality control and security screening).

## 2. Intensity correlations in in-line imaging

As shown in Fig.1, let a partially-coherent light source be located in the plane $z = -R_1$ perpendicular to the optic axis $z$; the source illuminates a thin semi-transparent object located immediately before the object plane $z = 0$, with the image, formed by the transmitted wave, recorded by a position-sensitive detector in the image (detector) plane $z = R_2$. The total source-to-detector distance is $R = R_1 + R_2$, and $M = R/R_1$ is the geometric magnification of the image relative to the object plane. For simplicity, we consider the case of two-dimensional in-line paraxial imaging in Cartesian coordinates $(x, z)$. The generalization of the obtained results to the corresponding full 3D picture expressed in coordinates $(x, y, z)$ is straightforward.

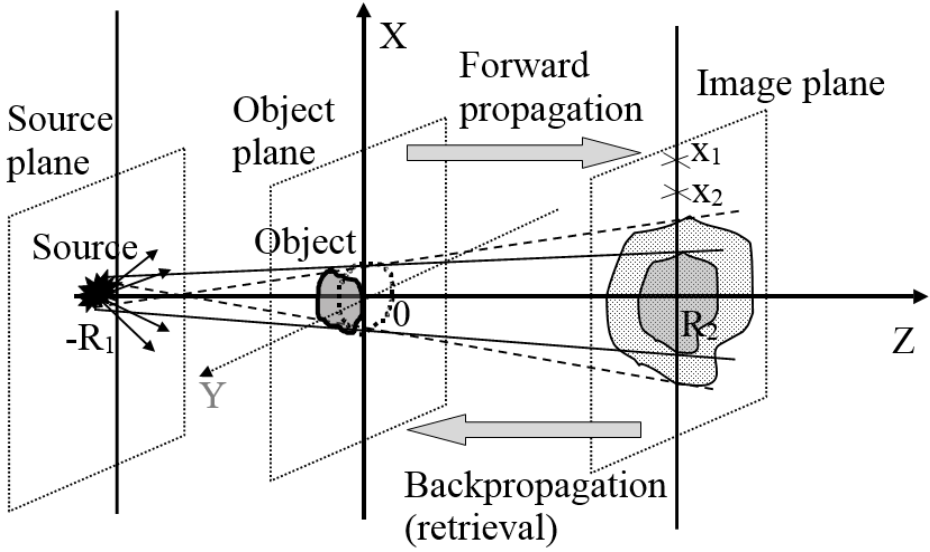

Fig. 1. Setup of (lensless) in-line free-space imaging considered in the present paper.

Let us consider the following form of covariance of registered light irradiance $I_{T,R}(x)$ at a pair of points $(x_1, x_2)$ in the image plane, $z = R_2$:

$$C_{T,R}(x_1, x_2) \equiv <I_{T,R}(x_1)I_{T,R}(x_2)> - <I_{T,R}(x_1)><I_{T,R}(x_2)>, \qquad (1)$$

where $I_{T,R}(x) = T^{-1}\int_{-T/2}^{T/2} |U_R(x,t)|^2 \, dt$ is the instantaneous light irradiance averaged over the exposure time $T$, $U_R(x,t)$ is the complex wave amplitude (viewed as a wide-sense stationary stochastic process) in the image plane and the angular brackets denote ensemble average. Note that $<I_{T,R}(x)> = I_R(x)$, where $I_R(x) \equiv \lim_{T \to \infty} I_{T,R}(x)$ is the time-averaged irradiance, because the ensemble average value of intensity is time-independent for wide-sense stationary processes.

It can be shown [13,14] that the covariance of the detected intensity distribution can typically be represented as a sum of two terms:

$$C_{T,R}(x_1, x_2) = C_{T,R}^{self}(x_1, x_2) + C_{T,R}^{det}(x_1, x_2), \quad (2)$$

where $C_{T,R}^{self}(x_1, x_2)$ is the covariance of "self-noise" of the electromagnetic field in the image plane and $C_{T,R}^{det}(x_1, x_2)$ is the covariance of the shot noise inherent to the photodection process. The latter ("detector") noise is usually spatially uncorrelated (excluding the effect of the detector PSF, see more on that below) and has Poisson statistics:

$$C_{T,R}^{det}(x_1, x_2) \cong T^{-1}\delta(x_1 - x_2)\int \eta^{-1}(\nu) S_R((x_1 + x_2)/2, \nu) d\nu, \quad (3)$$

where $S_R(x)$ is the spectral density of light in the image plane and $\eta$ is a constant characterizing the detector efficiency [14], such that $\bar{n}_\rho(x,\nu) = \eta(\nu) T S_R(x,\nu)$ is the average number of photons registered over the exposure time $T$ per unit length and per unit frequency range. In eq.(3), the assumed continuity of $S_R(x,\nu)$ implies that $\delta(x_1 - x_2) S_R(x_1, \nu) = \delta(x_1 - x_2) S_R((x_1 + x_2)/2, \nu)$. Note that the covariance in eqs.(1)-(3) is defined in terms of the correlation of light irradiance at different spatial positions, a situation which has obvious parallels, for example, with the Hanbury Brown - Twiss experiments [13-18] and with recent studies in X-ray ghost imaging [19-22]. Note also that, due to our choice of variables in eq.(1), the dimensionality of $C_{T,R}(x_1, x_2)$, as well as both of its components on the right-hand side of eq.(2), is $J^2/m^2/s^2$. The dimensionality of the detector efficiency constant, $\eta$, is $J^{-1}$, and it effectively converts the incident spectral energy density into the number density of detected photons (of the corresponding frequency).

The "self" noise depends on the properties of the light source, interaction with the sample and free-space propagation. Here we assume the optical properties of the sample to be static and deterministic. In this case, when the source has circular Gaussian statistics [13], one can use the Gaussian moments theorem to derive the following expression for the covariance of the self-noise [23]:

$$C_{T,R}^{self}(x_1, x_2) = \iint W_R(x_1, x_2, \nu_1) W_R^*(x_1, x_2, \nu_2) \operatorname{sinc}^2[\pi(\nu_1 - \nu_2)T] d\nu_1 d\nu_2, \quad (4)$$

where $W_R(x_1, x_2, \nu)$ is the cross-spectral density of light [14] in the image plane. In the case in which the coherence time of the radiation is much shorter than the exposure time $T$, the previous expression can be simplified [23]:

$$C_{T,R}^{self}(x_1, x_2) \cong T^{-1} \int |W_R(x_1, x_2, \nu)|^2 \, d\nu. \tag{5}$$

In the special case, when the light is cross-spectrally pure [14,24], i.e. when $W_R(x_1, x_2, \nu) = J_R(x_1, x_2)s(\nu)$, with $s(\nu) \geq 0$ and $\int s(\nu)d\nu = 1$, eq.(5) simplifies to

$$C_{T,R}^{self}(x_1, x_2) = (T_c / T) |J_R(x_1, x_2)|^2, \tag{6}$$

where $T_c = \int s^2(\nu)d\nu$ is the coherence time and $J_R(x_1, x_2)$ is the mutual intensity [13,14, 25]. The coherence time $T_c$ can also be introduced in the case of the more general eq.(5) (i.e. without the assumption of cross-spectral purity) in a less explicit manner as an inverse of the spectral width. One can then introduce the dimensionless integration variable $\nu' = T_c^{-1}\nu$ in eq.(5), which will lead to the multiplicative factor $T_c / T$ appearing just as in eq.(6). Note that equations (5) and (6) are derived under conditions that imply $T_c / T \ll 1$, which is true for almost all imaging experiments, except, perhaps, those employing the highest-quality single-mode lasers as light sources [23]. In cases with $T_c / T \ll 1$, according to eq.(2), the variance of the detected intensity will be typically dominated by the detector shot noise (see also eq.(21) below). However, as this noise is uncorrelated beyond the width of the detector PSF, the covariance may still be dominated by the self-noise term for pairs of points $x_1, x_2$ separated by a distance larger than the width of the detector PSF. Note that we are not assuming cross-spectrally pure light for the bulk of the remainder of this paper.

Let us now consider the general effect of convolution of the intensity covariance with some point-spread function (PSF), $P(x)$: $C_{P,T,R}(x_1, x_2) = \iint C_{T,R}(x_1 - x_1', x_2 - x_2')P(x_1')P(x_2')dx_1'dx_2'$. This can correspond, for example, to detector PSF or other post-detection filtering of the registered intensities. We assume that the PSF is normalized in the usual way: $\int P(x)dx = 1$. We also assume for simplicity that both the square modulus of the cross-spectral density, $|W_R(x_1, x_2, \nu)|^2$, and the spectral density, $S_R(x, \nu)$, vary slowly over distances comparable with the width of the PSF, $P(x)$. This allows us to neglect the effect of image blurring due to the convolution with the PSF, which is rather straightforward to take into account, but would have complicated many subsequent expressions without altering the key conclusions. As a result, the only effect of the PSF that we do take into account here is the corresponding change of the transverse correlation length of the shot noise. Equations (2), (3) and (5) then imply that

$$C_{P,T,R}(x_1, x_2) = C_{P,T,R}^{self}(x_1, x_2) + C_{P,T,R}^{det}(x_1, x_2)$$
$$\cong T^{-1} \int [|W_R(x_1, x_2, \nu)|^2 + P_2(x_2 - x_1)\eta^{-1}(\nu) S_R((x_1 + x_2)/2, \nu)] d\nu, \quad (7)$$

where $P_2(x) = \int P(x + x')P(x')dx'$ is the autocorrelation of the PSF.

In view of eq.(7), in order to calculate the free-space propagation of intensity covariance, it is sufficient to calculate the free-space propagation of the cross-spectral density. The corresponding derivation is presented separately in the Appendix, where we have used an approach similar to that in Refs.26, 27 in order to derive an expression for free-space propagation of cross-spectral density in a general setting corresponding to Fig.1. For simplicity, here we only consider the case of plane wave illumination. It is shown in the Appendix that in this case $|W_R(x_1, x_2, \nu)|^2 = S_R(x_1, \nu)S_R(x_2, \nu)$, where $S_R(x, \nu) = |Q_R(x, \nu)|^2$ and $Q_R(x, \nu)$ is the Fresnel diffraction integral of the function $Q_0(x, \nu) = S_{in}^{1/2}(x, \nu) q(x, \nu) \exp[ik\psi_{in}(x, \nu) + i\varphi(x, \nu)]$, which represents the transmission function of the imaged object, $Q(x, \nu) = q(x, \nu) \exp[i\varphi(x, \nu)]$, modulated by the incident illumination (see eq.(A2) and (A8) in the Appendix). Consequently, eq.(7) leads to the following expression for the intensity covariance in the image plane:

$$C_{P,T,R}(x_1, x_2) \cong T^{-1} \int [S_R(x_1, \nu) S_R(x_2, \nu) + P_2(x_2 - x_1)\eta^{-1}(\nu) S_R((x_1 + x_2)/2, \nu)] d\nu. \quad (8)$$

In the case of cross-spectrally pure light, $S_R(x, \nu) = I_R(x) s(\nu)$, the last result becomes

$$C_{P,T,R}(x_1, x_2) \cong (T_c/T) I_R(x_1) I_R(x_2) + T^{-1} \eta^{-1} P_2(x_2 - x_1) I_R((x_1 + x_2)/2), \quad (9)$$

where we have also assumed for simplicity that the detector efficiency is the same for all frequencies, $\eta(\nu) = \eta$, and we used the facts that, by definition, $\int s(\nu) d\nu = 1$ and $\int s^2(\nu) d\nu = T_c$. Equation (9) has a very simple and physically transparent form: here the intensity covariance is a sum of a Gaussian self-noise term and a Poisson shot-noise term which is correlated only within the width of the autocorrelation function of the detector PSF.

The following equation for the intensity variance, $V_{P,T,R}(x) \equiv C_{P,T,R}(x,x)$, in the spatially coherent and cross-spectrally pure case follows from eq.(9) at $x_1 = x_2$:

$$V_{P,T,R}(x) \cong (T_c/T)I_R^2(x) + T^{-1}\eta^{-1}(\Delta_2 x)^{-1} I_R(x), \quad (10)$$

where $\Delta_2 x = 1/P_2(0) = 1/\int P^2(x)dx$ represents an alternative measure of the spatial resolution of the imaging system, i.e. of the width of the detector PSF [5]. Note, for example, that if $P(x)$ is a Gaussian with standard deviation $\sigma$, then $\Delta_2 x = 2\pi^{1/2}\sigma$; if $P(x)$ is a Lorentzian (Cauchy) distribution with FWHM $2\sigma$, then $\Delta_2 x = 2\pi\sigma$; if $P(x)$ is a rectangular function with side length $2\sigma$, then $\Delta_2 x = 2\sigma$, etc. Here again the first (i.e. self noise) term has behaviour indicative of Gaussian statistics, with the variance being proportional to the square of the intensity, while the second (i.e. detector noise) term clearly corresponds to the usual statistics of shot noise, with the variance being proportional to the average intensity and inversely proportional to the width of the detector PSF. The second term in eq.(10) is typically dominant, as long as $T_c/T << 1$. This fact is particularly easy to see after multiplying both sides of eq.(10) by $T^2\eta^2(\Delta_2 x)^2$: after that, the first term on the right-hand side of eq.(10) becomes $(T_c/T)[\bar{n}(x)]^2$, while the second term is equal to $\bar{n}(x)$, where $\bar{n}(x) = \eta(\Delta_2 x)TI_R(x)$ is the average number of photons registered within an area of the size $(\Delta_2 x)$ around point $x$ during the exposure time $T$. Note, for example, that for X-rays with wavelength $\lambda = 0.1$ nm and monochromaticity $\Delta\lambda/\lambda \sim 10^{-4}$, the coherence time is $T_c \sim 1/\Delta\nu = \lambda^2/(c\Delta\lambda) \cong 3\times 10^{-15}$ s. If the exposure time is a few milliseconds, then the number of photons registered in a pixel during this exposure needs to be of the order of $10^6$ or higher for the first term in eq.(10) to become larger than the second one.

Using the results from [23], it is easy to obtain also an analogue of eq.(8) for the temporally coherent case, i.e. for the case when the exposure time $T$ is much shorter than the coherence time $T_c$ ($T_c/T >> 1$):

$$C_{P,T,R}(x_1, x_2) \cong |\int \exp\{ik/(2R')(x_1^2 - x_2^2)\}Q_R^*(x_1,\nu)Q_R(x_2,\nu)d\nu|^2 \\ + T^{-1}P_2(x_2-x_1)\int \eta^{-1}(\nu)S_R((x_1+x_2)/2,\nu)d\nu. \quad (11)$$

Note that in this case, the first (i.e. self) term is no longer divided by the exposure time (see [23] for details). This regime was considered in a recent publication [28] in the case of far-field imaging (diffraction). It requires extremely short exposure times or light pulse durations. In the example considered above, with X-rays of wavelength $\lambda = 0.1$ nm and monochromaticity $\Delta\lambda/\lambda \sim 10^{-4}$, the pulse duration has to be shorter than a femtosecond in order for the regime described by eq.(11) to apply.

**3. Implications for TIE-Hom imaging in near-Fresnel region**

Under the conditions of near-Fresnel imaging, one could use the well-known Transport of Intensity approximation (see e.g. [8, 26]) in order to further simplify eq.(10) and better understand its behaviour upon free-space propagation. This equation, namely the continuity equation associated with the parabolic equation obeyed by paraxial scalar electromagnetic waves, may be applied in its finite-difference form [8], if one assumes the object-to-detector propagation distance to be sufficiently small for the intensity at each transverse position to evolve linearly with propagation distance.

Here we consider plane wave illumination for simplicity, but the results can be generalized to the more general spatially partially-coherent case. Furthermore, in order to demonstrate the key points in the simplest possible way, we assume that the incident amplitude and the complex transmission function of the object satisfy the monomorphous conditions and that the "homogeneous TIE" approximation can be used [8], defined by

$$S_R(x,\nu) \cong [1 - \sigma^2 \nabla_x^2] S_0(x,\nu), \qquad (12)$$

where $S_0(x,\nu) = S_{in}(\nu)|q(x,\nu)|^2$ is the transmitted spectral density in the object plane, $\sigma^2 = \gamma R'/k$ and $[\gamma(\nu)/2]\ln S_0(x,\nu) = k\psi_{in}(\nu) + \varphi(x,\nu)$ according to the assumption of "monomorphicity". An important special case of monomorphicity is a sample composed of a single material of variable density, although the class of monomorphous objects is broader than this. In the simplest case, one typically considers spectrally-uniform illumination with $S_{in}(\nu) \cong 1$ and $\psi_{in}(\nu) \cong 0$, and a monomorphous (also called homogeneous) object with $\varphi(x,\nu) = \gamma(\nu)\ln q(x,\nu)$, with $\gamma(\nu) = \delta(\nu)/\beta(\nu)$ and $1 - \delta(\nu) + i\beta(\nu)$ being the complex refractive index of the object. Substituting eq.(12) into eq.(8) and setting $x_1 = x_2 = x$, we obtain

$$V_{P,T,R}(x) \cong T^{-1}\int\{[(1-\sigma^2\nabla_x^2)S_0(x,\nu)]^2 + \eta^{-1}(\nu)(\Delta_2 x)^{-1}(1-\sigma^2\nabla_x^2)S_0(x,\nu)\}d\nu \qquad (13)$$

for the propagation of (noise) variance. Note that according to the validity conditions of the TIE-Hom model (eq.(12)), the "propagation contrast" must be weak [8], in the sense that

$$\left|\frac{\sigma^2 \nabla^2 S_0(x,\nu)}{S_0(x,\nu)}\right| = \left|\frac{S_R(x,\nu)}{S_0(x,\nu)} - 1\right| \ll 1. \tag{14}$$

This is particularly true within flat regions of the image, where the noise level is typically measured (by assuming the spatial ergodicity and replacing ensemble average by the spatial average over a uniform area of an image [9]). In a spatially uniform area of an image, the Laplacian of the spectral density will be very small. This allows one to neglect the terms involving the Laplacian in eq.(13), leading to the equation

$$V_{P,T,R}(x) \cong V_{P,T,0}(x). \tag{15}$$

Equation (15) shows that the variance of image noise remains approximately constant during free-space propagation within the near-Fresnel region. The same will obviously be true for the SNR in the image plane

$$SNR_R(x) = \frac{\bar{I}_R(x)}{V_{P,T,R}^{1/2}(x)} \cong \frac{\bar{I}_0(x)}{V_{P,T,0}^{1/2}(x)}, \tag{16}$$

where the corresponding "signals" in the image and object planes, $\bar{I}_R(x)$ and $\bar{I}_0(x)$ respectively, are equal to the ensemble-averaged intensities $I_R(x) = \int S_R(x,\nu)d\nu$ and $I_0(x) = \int S_0(x,\nu)d\nu$, averaged over the same uniform image region where the noise is evaluated.

It has been demonstrated in a number of recent publications [10, 29-31], that the application of the transformation inverse to TIE-Hom, eq.(12), may allow one to significantly reduce the noise variance in the image and boost SNR. As stated previously, an important motivation for the present study is provided by the fact that SNR boosts of two orders of magnitude are possible, enabling four orders of magnitude reduction in both dose and acquisition time. The corresponding transform, $S_{0,ret}(x,\nu) = [1 - \sigma^2 \nabla_x^2]^{-1} S_R(x,\nu)$ (which is usually called TIE-Hom retrieval, or the Paganin method [8]), can be expressed as a convolution with the suitable filter function

$$S_{0,ret}(x,\nu) = \int P_\sigma(x - x') S_R(x',\nu) dx', \tag{17}$$

with $P_\sigma(x) = \int \exp(-i2\pi x\xi)(1+4\pi^2\sigma^2\xi^2)^{-1}d\xi = (2\sigma)^{-1}\exp(-\sigma^{-1}|x|)$. Note that $\int P_\sigma(x)dx = 1$. Similarly to the derivation of eq.(7) above, it is possible to show (see also [5]) that if the width of this filter function is much larger than the width of the detector PSF, but much smaller than the characteristic variation length of the function $S_R(x,\nu)$, then the noise variance satisfies

$$V_{P,T,0}^{ret}(x) = V_{P_\sigma,P,T,R}(x) = \iint C_{P,T,R}(x-x_1', x-x_2')P_\sigma(x_1')P_\sigma(x_2')dx_1'dx_2' \\ \cong V_{P,T,R}^{self}(x) + [\|P_\sigma\|_2^2 / P_2(0)]V_{P,T,R}^{det}(x) \cong [(\Delta_2 x)/(4\sigma)]V_{P,T,R}(x), \quad (18)$$

where $[P_2(0)]^{-1} = \Delta_2 x$ as above, and $\|P_\sigma\|_2^2 = \int P_\sigma^2(x)dx = (2\sigma)^{-2}\int \exp(-2\sigma^{-1}|x|)dx = (4\sigma)^{-1}$. In the derivation of eq.(18), we have also taken into account the facts that $\int P_\sigma(x)dx = 1$ and $V_{P,T,R}^{self}(x) << V_{P,T,R}^{det}(x)$. According to eq.(18), the variance of image noise reduces upon TIE-Hom retrieval in proportion to the ratio of the corresponding spatial resolutions $4\sigma/(\Delta_2 x) = 4\{\gamma R'/[(\Delta_2 x)^2 k]\}^{1/2} >> 1$. This ratio is large because we assumed that the width of the filter function is much larger than that of the detector PSF. Note also that the "signal" $\bar{I}_R(x)$ does not change under the convolution with the filter function, as we assumed that the spectral density spatially varies much more slowly than the filter function. Therefore, the SNR increases in proportion to the factor $[4\sigma/(\Delta_2 x)]^{1/2}$ upon the TIE-Hom retrieval. In the case of more conventional 2D images, the corresponding factor is even larger, being equal to $2\pi^{1/2}\sigma/(\Delta_2 x)$ [9]. This is fully consistent with the noise-resolution uncertainty principle [5,6].

There is generally nothing special about the reduction of image noise and increase of SNR as a result of low-pass filtering of an image in accordance with eq.(18). Indeed, any other low-pass filter in place of the TIE-Hom retrieval would achieve a similar result. The fact that noise correlation length increases and noise variance decreases as a result of low-pass filtering is well known [32]. What is truly remarkable about the TIE-Hom retrieval procedure, when used in combination with free-space propagation, is that it manages to reduce image noise without making the spatial resolution worse, compared with the "contact" image collected in the object plane $z = 0$ under equivalent conditions. Indeed, it is known that any linear filtering operation, which can be expressed by a convolution equation of the same type as eq.(17), increases the width of the PSF of the imaging setup in the same proportion as it decreases the image noise [5]. In the case of TIE-Hom imaging with plane-wave illumination, however, the spatial resolution is first improved during free-space propagation from the object to the image plane, and then it is returned to the object-plane level during the TIE-Hom retrieval. Therefore, there is no overall change in the spatial resolution as the combined result of the forward propagation and the retrieval. At the same time, as shown by eqs.(15)-(16), the image noise and the SNR do not change during free-space propagation, while the noise variance decreases and the SNR increases during TIE-Hom retrieval, according to eq.(18), leading to the net noise decrease and SNR increase as the combined result of the forward propagation and the retrieval.

Let us make the above important point more quantitative. Let the spectral density distribution in the object plane be represented as $S_0(x,\nu) = (S_0^{id} * P)(x,\nu)$, where $S_0^{id}(x,\nu)$ is an "ideal" (sharp) image, $P(x)$ is the energy-independent PSF of the detector (with unit integral norm and zero expectation value) and the asterisk sign denotes spatial convolution. Similarly, $S_R(x,\nu) = (S_R^{id} * P)(x,\nu)$, as we assume the detector PSF to be the same in both planes. The spatial resolution of an imaging system can be defined as the standard deviation of the PSF which is convolved with an ideal sharp image

$$(\Delta x)[P] = \left(4\pi \int x^2 P(x)dx\right)^{1/2}, \tag{19}$$

where the normalization constant $4\pi$ is chosen for consistency with our recent publications [5,6] (this makes the width of a Gaussian PSF equal to $2\pi^{1/2}\sigma$ and the width of a rectangular PSF equal to $(\pi/3)^{1/2}\sigma$, where $\sigma$ is the standard deviation of the Gaussian and the side length of the rectangle, respectively). As $S_R(x,\nu) = [1-\sigma^2\nabla_x^2]S_0(x,\nu) = (S_0^{id} * [1-\sigma^2\nabla_x^2]P)(x,\nu)$, it is then straightforward to verify that the forward TIE-Hom propagation, eq.(12), improves the spatial resolution in the sense that the effective variance of the PSF is decreased

$$(\Delta x)_R^2[P] = (\Delta x)_0^2[(1-\sigma^2\nabla_x^2)P] = (\Delta x)_0^2[P] - 8\pi\sigma^2, \tag{20}$$

as $4\pi \int x^2(1-\sigma^2\nabla_x^2)P(x)dx = (\Delta x)_0^2[P] - 4\pi\sigma^2 \int (\nabla_x^2 x^2)P(x)dx = (\Delta x)_0^2[P] - 8\pi\sigma^2$. Note that the spatial resolution in eq.(20) is referenced to the object plane. Equation (12), which describes the Fresnel diffraction of the spectral density of a thin monomorphous object under the assumption of large Fresnel number, is mathematically identical in form to the unsharp-mask image sharpening algorithm in which the unsharp mask is approximated by the negative Laplacian of a greyscale image. More generally, eq.(12) represents a generic second-order local approximation to image deconvolution with a PSF having the second central integral moment equal to $2\sigma^2$ [33]. TIE-Hom retrieval has exactly the opposite effect to such image sharpening, so that

$$(\Delta x)_{0,ret}^2[P] = (\Delta x)_R^2[(1-\sigma^2\nabla_x^2)^{-1}P] = (\Delta x)_R^2[P] + 8\pi\sigma^2 = (\Delta x)_0^2[P]. \tag{21}$$

Therefore, the combined effect of the free-space propagation and phase retrieval leaves the spatial resolution in the object plane unchanged.

As a result, the free-space propagation followed by the corresponding TIE-Hom retrieval can significantly improve the ratio of SNR to spatial resolution, i.e. improve the "intrinsic imaging system quality" [6], according to

$$\frac{SNR_{0,ret}(x)}{(\Delta x)_{0,ret}} \cong \left(\frac{4\sigma}{h}\right)^{1/2} \frac{SNR_0(x)}{(\Delta x)_0}. \qquad (22)$$

It has been demonstrated both theoretically [9] and experimentally [10] that the improvement factor can reach several hundreds (in the 3D case) without violating the TIE-Hom validity conditions. This in turn enables a reduction in the radiation dose by factors in the tens of thousands [10]. Therefore, the method has a potential to revolutionise biomedical X-ray imaging, in particular [10, 29-31].

The demonstrated increase in the ratio of SNR to spatial resolution is even more remarkable given that we have recently proven [5] that this ratio must remain essentially constant in linear (with respect to intensity) shift-invariant (LSI) transformations of the imaging system. As both the TIE-Hom forward propagation, eq.(12), and the TIE-Hom retrieval, eq.(17), are LSI transformations with the normalized PSFs, there appears to be a contradiction between the two results. Note however that if eq.(12) was applied to the spectral density of a detected image in the object plane (which would include the shot noise), followed by the TIE-Hom retrieval, eq.(17), the net result would indeed be no change in both the spatial resolution and the SNR. The gain in the SNR to spatial resolution ratio demonstrated by eq.(22) is achieved because the shot noise is only added to the measured spectral density during the image acquisition in the image plane (this noise is absent in the forward propagation stage), after which it is suppressed by the TIE-Hom phase retrieval [9]. It is also important that in the considered case the shot noise is dominant. Note that the self noise behaves differently. If considered in the zero-order approximation (i.e. neglecting the Laplacian term in the TIE-Hom), the self noise remains constant both upon forward propagation and at the retrieval stage (see eqs.(15) and (18)). If considered in the "first-order" approximation, in which the Laplacian terms are taken into account, the self-noise increases during the forward propagation, followed by the equivalent decrease during the retrieval stage, again giving zero net effect.

## 4. Optical deconvolution and other generalizations

Here we give an indication of how a broader class of imaging systems, than those we have hitherto considered, might exhibit "unreasonable" effectiveness similar to that of TIE-Hom. To this end, consider imaging systems of the form

$$S_R(x,v) \cong P_h(x) * [1 - \sum_{j=1}^{\infty} \alpha_j \partial_x^j] S_0(x,v), \qquad (23)$$

which is a generalization of eq.(23), where $P_h(x)$ is a low-pass filter which cuts all spatial frequencies above a certain threshold, i.e. $\hat{P}_h(\xi) = 0$ for $|\xi| > 1/h$, and all of the coefficients $\alpha_j$ are real. In real setups, the role of the low-pass filter $P_h(x)$ can be played by the PSF of the detector, the aperture of a lens or by other components of the imaging system limiting the range of spatial frequencies from above. We further assume that

(a) $\alpha_j$ is non-negative for $j = 2, 6, 10 \ldots$ and non-positive for $j = 4, 8, 12 \ldots$,

(b) at least one of the $\alpha_j$ coefficients is non-zero, and

(c) a condition generalising eq.(14) holds, namely that

$$\sum_{j=1}^{\infty} |\alpha_j| h^{-j} < 1. \qquad (24)$$

These stated conditions ensure that the operator in squared brackets in eq.(23) amplifies the lower frequencies (up to the cut-off frequency $1/h$) of the input (the spectral density distribution in the object plane), and it also has a reciprocal that is a compact operator, ensuring that no information about the image is lost upon the action of the corresponding optical transformation. Special cases of eq.(23), besides eq. (12) above, include:

(1) analyser-based X-ray phase-contrast imaging of monomorphous objects under the approximation that the analyser-crystal rocking curve may be linearised as function of spatial frequency, in regions of Fourier space where the Fourier representation of the illuminating field is non-negligible [34-36],

(2) phase-contrast imaging of monomorphous objects using combined analyser-crystal and propagation based phase contrast [34,37] or edge-illumination based phase contrast [38], and

(3) phase-contrast imaging of weak monomorphous objects using arbitrary coherent LSI imaging systems [39], whose aberration coefficients (e.g. coma, astigmatism, defocus, spherical aberration etc.) are chosen to satisfy the auxiliary conditions on $\alpha_j$ that are specified above.

Let us consider one further generalization of the above approach. It is easy to show that a convolution of intensity distribution with a PSF $P(x)$ can be expressed in the form of Taylor series, i.e.

$$I_P(x) = \int P(x-x')I(x')dx = [1 + \sum_{j=1}^{\infty} p_j \partial_x^j]I(x), \text{ with } p_j = \frac{(-1)^j}{j!}\int x^j P(x)dx. \quad (25)$$

It has been shown in Refs.33,40 that, under certain conditions on the PSF function $P(x)$, the corresponding deconvolution can be also expressed in the form of a convolution or as a Taylor series:

$$I(x) = \int A(x-x')I_P(x')dx = [1 + \sum_{j=1}^{\infty} a_j \partial_x^j]I_P(x), \text{ with } a_j = \frac{(-1)^j}{j!}\int x^j A(x)dx, \quad (26)$$

with the coefficients (integral moments) satisfying the following iterative relations

$$a_j = -p_j - \sum_{i=1}^{j-1} a_i p_{j-i}, \; j = 1, 2, 3.... \quad (27)$$

If the function $P(x)$ is a typical "well-behaved" PSF, e.g. a smooth, symmetric, non-negative function with a unit integral and zero expectation, then we will have from eq.(25) that $p_j = 0$ for odd $j$, and $p_j > 0$ for even $j$. In that case eq.(26) implies that $a_j = 0$ for odd $j$, and $a_j$ are likely to be negative or have alternating signs for even $j > 0$, as eq.(27) implies that $a_2 = -p_2$ (as in eq.(12)), $a_4 = -p_4 + p_2^2$, $a_6 = -p_6 + 2p_4 p_2 - p_2^3$, etc. Therefore, the "deconvolution kernels" $A(x)$ are likely to be rapidly oscillating and have negative, as well as positive, values, unlike the "well-behaved" convolution kernels $P(x)$ in eq.(25).

If an optical setup can implement the LSI transformation of the input intensity described by eqs.(26)-(27), it will obviously perform an exact deconvolution, in the sense that it will fully reverse the effect of the convolution with the "original" PSF, $P(x)$. However, the notion of optical deconvolution can be meaningfully defined in a broader sense. Let us postulate that a (partial) deconvolution is an LSI transformation that reduces the width of the system's PSF, or, more generally, which increases the width of the essential support of the Fourier transform of the intensity distribution. If the width is defined according to the standard deviation of the PSF, as in eq.(19), then, because $(\Delta x)^2[A*P] = (\Delta x)^2[A] + (\Delta x)^2[P] = 8\pi(a_2 + p_2)$, provided that both PSFs have unit integrals and zero expectations, the LSI transformation in eq.(26) will be a deconvolution if and only if $a_2 < 0$. Note that for such a partial deconvolution eq.(27) is not assumed to hold, and no reference to the "original" PSF, $P(x)$, is needed. Partial deconvolution is performed, for example, by the PSF $A(x) = [1 - \sigma^2 \partial_x^2]\delta(x)$, for which $(I*A)(x) = [1 - \sigma^2 \partial_x^2]I(x)$ and, hence, $a_2 = -\sigma^2$. Such examples are not limited to differential operators: consider e.g. a smooth function $A(x;\sigma,b) = [b\sigma^2 - (b-1)x^2]/(\sigma^3\sqrt{2\pi})\exp[-x^2/(2\sigma^2)]$. It is straightforward to verify that $\int A(x;\sigma,b)dx = 1$ for any $b$ and $\sigma \neq 0$, and $a_2 = \sigma^2(3/2 - b)$. Therefore, if $b > 3/2$, then $a_2 < 0$ and the LSI transformation eq.(26) with such a function $A(x;\sigma,b)$ performs a (partial) deconvolution.

An optical setup implementing eq.(26) prior to image detection, with a function $A(x)$ that has a negative second integral moment, will improve the effective spatial resolution and, if the noise is dominated by the photodetection shot noise, it will not change the average noise level in the output images. This property is in stark contrast with any numerical (post-detection) deconvolution, which typically increases the image noise. Numerical deconvolutions are different from optical ones in that numerical deconvolutions are applied to detected images which already contain photon shot noise (and possibly other detector noises), and the latter noise is inevitably amplified during the deconvolution operation. On the other hand, optical deconvolutions are applied prior to image detection, when the shot noise has not yet been generated. Therefore, any optical deconvolution system will have a similar "unreasonable" effectiveness to the TIE-Hom imaging described above. We have explained that the key to this effectiveness is in the "forward" process that improves the spatial resolution without increasing the noise. The subsequent numerical low-pass filtering, which can decrease the noise level and proportionally worsen the spatial resolution, does not really add anything special to the overall process. Some examples of optical deconvolution setups can be found in Ref.40. The existence of further such optical setups, a thorough exploration of which would form an interesting avenue for future research, gives added impetus to the work presented here.

## 5. Conclusions

We have derived generic formulae for free-space propagation of the covariance of light irradiance transmitted through a weakly scattering object illuminated by a partially coherent beam. We showed that the two components of this covariance, corresponding to Gaussian self noise (due to the statistical properties of the light source, the transmission and the free-space propagation) and to Poisson shot noise (due to the photodetection), respectively, both propagate in a rather straightforward manner determined primarily by the corresponding propagation laws of the cross-spectral density distribution $W_{T,R}(x_1, x_2, \nu)$. In the temporally incoherent case (which corresponds to almost all present-day imaging experiments with electromagnetic radiation), the shot noise term is dominant in the expression for the variance, obtained by setting $x_1 = x_2$, while the self noise term of the covariance can become dominant for pairs of points separated by distances larger than the diameter of the detector PSF. In the temporally coherent case, which may be realised e.g. in some experiments at X-ray free electron laser sources, the self term of the covariance may become dominant even at $x_1 = x_2$. This implies a possibility for intensity correlation imaging in the future using such sources [28].

An interesting analogy exists between the behaviour of image self-noise and the effect of source size on the spatial resolution of an in-line imaging system: the effective source size also increases upon forward propagation (in the case of non-trivial geometric magnification, $M = (R_1 + R_2)/R_1 > 1$), and then it reduces when the spatial resolution in the image is referred back to the object plane, with the net effect corresponding to a factor of $(M-1)/M$ [11]. On the other hand, the component of the system's spatial resolution which is due to the width of the detector PSF, is the same in the object and image planes, when measured in the respective planes. It is multiplied by the factor $1/M$ when the resolution is referred from the image plane back to the object plane. This behaviour is somewhat similar to that of the photodetection shot noise, which is also the same in the image and object planes (apart from the possible trivial factor $1/M$ reflecting the reduction of photon density in the case of quasi-spherical wave propagation, see Appendix), and is decreased during TIE-Hom retrieval (backpropagation). It may be interesting to investigate this analogy in more detail in the future.

In the present paper, we have used the behaviour of the image noise and spatial resolution to find a quantitative explanation of the "unreasonable" effectiveness of the TIE-Hom imaging. This explanation finally reconciles the apparent contradiction of this behaviour with the noise-resolution uncertainty principle [6]. It also provides a firm theoretical foundation for the SNR-boosting and resolution-preserving properties of the TIE-Hom method [8] of X-ray phase-contrast imaging that has been rapidly gaining popularity in recent years, particularly in view of its promising applications to medical and biomedical imaging [10,29-31]. Some indications have been also given, of how some of these results generalise to other imaging systems which can be shown to perform optical deconvolutions.

**Acknowledgements**

T.E.G. is grateful to Dr. Marcus Kitchen for useful discussions relevant to the present paper.

**Appendix**

In this appendix we derive an equation for free-space propagation of the cross-spectral density distribution in the setup shown in Fig.1.

Let the incident cross-spectral density in the object plane $z = 0$ be defined according to a generalized Schell model [26]

$$W_{in}(x_1, x_2, \nu) = S_{in}^{1/2}(x_1, \nu) S_{in}^{1/2}(x_2, \nu) \exp\{ik[\psi_{in}(x_2, \nu) - \psi_{in}(x_1, \nu)]\} \\ \times \exp[ik(x_2^2 - x_1^2)/(2R_1)] g_{in}(x_2 - x_1, \nu), \qquad (A1)$$

where $\tilde{g}_{in}(x_1, x_2, \nu) = \exp\{ik[\psi_{in}(x_2, \nu) - \psi_{in}(x_1, \nu)]\} \exp[ik(x_2^2 - x_1^2)/(2R_1)] g_{in}(x_2 - x_1, \nu)$ is the spectral degree of coherence and $S_{in}(x, \nu)$ is the incident spectral density. This model corresponds to a small polychromatic source located in the plane $z = -R_1$. The model includes as special cases the case of a point source, an incident plane wave and some typical partially coherent illumination produced at synchrotron beamlines [26].

Assuming a static thin object and the projection approximation for the transmission of light through the object, the transmitted cross-spectral density in the object plane can be expressed as

$$W_0(x_1, x_2, \nu) = W_{in}(x_1, x_2, \nu) Q^*(x_1, \nu) Q(x_2, \nu) \\ = Q_0^*(x_1, \nu) Q_0(x_2, \nu) \exp[ik(x_2^2 - x_1^2)/(2R_1)] g_{in}(x_2 - x_1, \nu), \qquad (A2)$$

where $Q(x, \nu) = q(x, \nu) \exp[i\varphi(x, \nu)]$ is the object transmission function and $Q_0(x, \nu) = S_{in}^{1/2}(x, \nu) q(x, \nu) \exp[ik\psi_{in}(x, \nu) + i\varphi(x, \nu)]$. Finally, assuming paraxial imaging conditions, the cross-spectral density in the image plane $z = R_2$ can be written via Fresnel diffraction integrals as

$$W_{M,R}(x_1, x_2, \nu) = \frac{1}{\lambda R_2} \iint \exp\{ik[(x_2 - x_2')^2 - (x_1 - x_1')^2]/(2R_2)\} W_0(x_1', x_2', \nu) dx_1' dx_2', \qquad (A3)$$

where the subscript index $M$ denotes the usual geometric magnification, $M = (R_1 + R_2)/R_1$, within the setup of Fig.1.

Let us define Hopkins' "effective source" [27] to be

$$g_{in}(x,\nu) = \int \exp(-ikxx'/R_1) S_n^{src}(x',\nu) dx', \qquad (A4)$$

where $S_n^{src}(x,\nu)$ is a spectral density distribution in the source plane normalized in such a way that $g_{in}(0,\nu) = \int S_n^{src}(x,\nu) dx = 1$. Substituting this into eqs.(A2) and (A3), we obtain

$$W_{M,R}(x_1, x_2, \nu) = \frac{1}{\lambda R_2} \exp\left[ik\left(\frac{x_2^2 - x_1^2}{2R_2}\right)\right] \iiint \exp\left\{ik\left[\frac{(x_1 x_1' - x_2 x_2')}{R_2}\right.\right.$$
$$\left.\left. + \frac{(x_1' - x_2')x''}{R_1} + \frac{(x_2')^2 - (x_1')^2}{2R'}\right]\right\} Q_0^*(x_1',\nu) Q_0(x_2',\nu) S_n^{src}(x'',\nu) dx_1' dx_2' dx'', \qquad (A5)$$

where $R' = R_1 R_2 / (R_1 + R_2) = R_2 / M$ is sometimes so-called the effective propagation (or defocus) distance [26]. The last equation can be re-written as

$$W_{M,R}(x_1, x_2, \nu) = \frac{1}{\lambda R_2} \exp\left[ik\left(\frac{x_2^2 - x_1^2}{2R_2}\right)\right] \iiint \exp\left\{\frac{ik}{2R'}\left[\tilde{x}_1^2 - \tilde{x}_2^2\right.\right.$$
$$\left.\left. + (\tilde{x}_2 - x_2')^2 - (\tilde{x}_1 - x_1')^2\right]\right\} Q_0^*(x_1',\nu) Q_0(x_2',\nu) S_n^{src}(x'',\nu) dx_1' dx_2' dx'', \qquad (A6)$$

where $\tilde{x}_{1(2)} = M^{-1} x_{1(2)} + (M-1)M^{-1} x''$. Let us introduce the function

$$W_R(x_1, x_2, \nu) = \exp[ik(x_1^2 - x_2^2)/(2R')] Q_R^*(x_1,\nu) Q_R(x_2,\nu), \qquad (A7)$$

where

$$Q_R(x,\nu) = \frac{\exp(ikR')}{i(\lambda R')^{1/2}} \int \exp[ik(x - x')^2 / (2R')] Q_0(x',\nu) dx' \qquad (A8)$$

is the Fresnel diffraction integral (under the condition of plane-wave illumination) of the complex-valued distribution $Q_0(x,\nu) = S_{in}^{1/2}(x,\nu) q(x,\nu) \exp[ik\psi_{in}(x,\nu) + i\varphi(x,\nu)]$ in the object plane. Then, apart from a multiplicative factor with modulus equal to 1, eq.(A6) represents a convolution of the "magnified" function $M^{-1} W_R(M^{-1}x_1, M^{-1}x_2, \nu)$ with the appropriately magnified and inverted normalised source spectral density distribution, $S_n^{src}(-(M-1)^{-1}Mx)$, of the form

$$W_{M,R}(x_1, x_2, \nu) = \exp\left[ik\left(\frac{x_2^2 - x_1^2}{2R_2}\right)\right]$$
$$\times M^{-1} \int W_R(M^{-1}x_1 - (M-1)M^{-1}x', M^{-1}x_2 - (M-1)M^{-1}x', \nu) S_n^{src}(-x', \nu) dx'. \tag{A9}$$

In particular, in the case of plane-wave illumination, when formally $R_1 = \infty$ and $M = 1$, we have

$$W_{1,R}(x_1, x_2, \nu) = \exp\left[ik\left(\frac{x_2^2 - x_1^2}{2R_2}\right)\right] W_R(x_1, x_2, \nu) = Q_R^*(x_1, \nu) Q_R(x_2, \nu), \tag{A10}$$

which means that in this case the cross-spectral density in the image plane is equal to the product $Q_R^*(x_1, \nu) Q_R(x_2, \nu)$ of propagated transmission coefficients (modulated by the incident illumination).

Another useful corollary of eq.(A9) is the corresponding equation for the spectral density, $S_{M,R}(x,\nu) = W_{M,R}(x,x,\nu)$, in the image plane:

$$S_{M,R}(x,\nu) = M^{-1} \int S_R(M^{-1}x - (M-1)M^{-1}x', \nu) S_n^{src}(-x', \nu) dx', \tag{A11}$$

which represents a convolution of the function $S_R(x,\nu)$ with the normalized source distribution (see also [26]).

We can now substitute the expression for the cross-spectral density in the image plane from eq.(A9) into the expression for the intensity covariance in eq.(7):

$$C_{P,T,R}(x_1,x_2) \cong T^{-1}M^{-2} \int | \int W_R(M^{-1}x_1 - (M-1)M^{-1}x', M^{-1}x_2 - (M-1)M^{-1}x_2', \nu)$$
$$\times S_n^{src}(-x',\nu)dx' |^2 \, d\nu + T^{-1}M^{-1}P_2(x_1 - x_2) \quad (A12)$$
$$\times \int \int \eta^{-1}(\nu) S_R(M^{-1}(x_1+x_2)/2 - (M-1)M^{-1}x', \nu) S_n^{src}(-x',\nu) dx' d\nu.$$

Equation (A12) has a simple structure: the function $W_R(x_1, x_2, \nu) = \exp[ik/(2R')(x_1^2 - x_2^2)]Q_R^*(x_1,\nu)Q_R(x_2,\nu)$ is convolved with the rescaled normalised spectral density of the effective source and then the square modulus of the result is integrated over all temporal frequencies; a similar, but simpler, convolution takes place for the shot noise component. Equation (A12) shows, in particular, that the self-noise generally decreases in proportion to the second power of the magnification, while the shot noise decreases only in proportion to the first power. Therefore, when the magnification factor increases, the shot-noise term becomes even more dominant, compared to self noise. This is to be expected, at least in retrospect: the more the image is magnified upon propagation, the more spatially dilute the photon density becomes. The ratio of the (typically Poissonian) shot noise to the (typically Gaussian) self noise therefore increases, on account of the different scaling behaviour of the two species of statistics, which implies that the fluctuations due to the former decrease (with increasing magnification) more slowly than those due to the latter.

In the case of plane-wave illumination, $M = 1$, the convolution with the normalised source spectral density distribution disappears and eq.(A12) becomes:

$$C_{P,T,R}(x_1,x_2) \cong T^{-1} \int [S_R(x_1,\nu)S_R(x_2,\nu) + P_2(x_1-x_2)\eta^{-1}(\nu)S_R((x_1+x_2)/2,\nu)]d\nu, \quad (A13)$$

where $S_R(x,\nu) = W_R(x,x,\nu) = |Q_R(x,\nu)|^2$ is the spectral density distribution in the image plane in the case of plane-wave illumination and $S_R(x_1,\nu)S_R(x_2,\nu) = |W_R(x_1,x_2,\nu)|^2$ according to eq.(A7).